# Controlled spacing of self-organized anodic TiO$_2$ nanotubes

Selda Ozkan,[1] Nhat Truong Nguyen,[1] Anca Mazare,[1] Patrik Schmuki[1, 2,*]

[1] *Department of Materials Science, Institute for Surface Science and Corrosion WW4-LKO, Friedrich-Alexander University, Martensstraße 7, D-91058 Erlangen, Germany*

[2] *Chemistry Department, Faculty of Sciences, King Abdulaziz University, 80203 Jeddah, Saudi Arabia Kingdom*

**\*** - Corresponding author:

   Prof. Dr. Patrik Schmuki:

Tel.: +49-9131-852-7575, Fax: +49-9131-852-7582, Email: schmuki@ww.uni-erlangen.de

**Abstract**

We report on how to grow and control self-organized TiO$_2$ nanotube arrays that show defined and regular gaps between individual nanotubes. For this we use electrochemical anodization of titanium in fluoride containing di-ethylene glycol (DEG) electrolytes, with variations in voltage and water content in the electrolyte. In these specific electrolytes, such nanotubes show a true spacing, i.e. nanotubes are spaced both at top and at bottom in regular intervals, this in contrast to classic nanotubes obtained in other organic electrolytes showing a close-packed organization. We identify critical parameters, that define the "region of existence" i.e. under which condition tube spacing occurs as well as the intertube distance, to be the voltage and the water content. Using these findings allows to grow tubes where diameter and spacing can even be independently controlled.

# 1. Introduction

Since the first reports on nanotubes or nanopores growth on Ti or Ti alloys by electrochemical anodization by Assefpour-Dezfuly [1] and later by Zwilling et al. [2], Gong et al. [3] and Beranek et al. [4], $TiO_2$ nanotubes have been explored for a wide range of applications and have, over the past decade, become one of the most investigated 1D nanostructures. Remarkable progress has been made in improving the morphology of $TiO_2$ nanotubes (NTs) and nowadays a high control over morphology (diameter, length, open top surfaces, smoothness of tube wall) has been achieved [5,6]. In turn, the control over morphology enabled considerable progress in the application of these well-ordered nanostructures, in a variety of domains such as electrical and photoelectrical applications, photoelectrochemical water splitting and biomedical coatings or drug release systems [6]. Current efforts for further progress for example target an even further increased ordering of the tubes [7] or creating fully defined single-walled tubes [8,9].

Generally, these NTs layers obtained in aqueous or organic electrolytes grow in a hexagonally close-packed configuration. This is not always apparent from top views as under many conditions nanotubes tend to grow in a bulb shape [6,10,11], but it is typically evident at the bottom growth front, close to substrate. Nevertheless, previous findings report that with certain electrolyte compositions, NTs growth with some spacing between individual tubes can be established. This particularly if the electrolyte is based, except for HF and $H_2O$, on DEG or di-methyl sulfoxide (DMSO) [12-16]. Over the years, few more electrolytes were reported to result in the growth of spaced or "loose-packed" NTs under specific anodization conditions e.g. ethylene glycol (EG), tri (tetra, poly)-ethylene glycol etc. [17-19].

Nevertheless, a systematic investigation of the key parameters that could be exploited to gain control over morphological features of such arrays, namely the intertube spacing, is missing.

In the present work, we evaluate and report the key anodization parameters that influence the occurrence (or not) of spaced tubes (as opposed to hexagonal close-packed tubes or other anodic oxide morphologies), and show how to control the morphology and spacing of these peculiar nanotube arrays.

## 2. Experimental

Ti foils (0.10mm, 99.6%, Advent, England) were degreased by ultrasonication (acetone, ethanol and distilled water) and dried in nitrogen stream. Prior to anodization, samples were pretreated (double anodization), anodized (in DEG + 4wt% HF (40%) + 0.3wt%$NH_4F$ + 1wt%$H_2O$), immersed in ethanol (1h) and dried. An electrochemical set-up (IMP-Series Jaissle Potentiostat) with a two-electrode configuration (Pt counter electrode, Ti working electrode) was used. Different voltages (5-50V) and different water contents were evaluated. Samples' morphology was investigated using a scanning electron microscope (SEM) Hitachi FE-SEM 4800. All geometrical properties (outer diameter, spacing and wall thickness) were measured from SEM images (Image J software), and are the average of 15 measurements.

## 3. Results and discussion

Figure 1 presents an example of spaced $TiO_2$ nanotubes obtained by anodization in the above described DEG electrolyte, at 30V for 4h. In this case, individual tubes with 1.2μm length and outer diameter of ≈150nm grow, but compared to the generally reported close-packed configuration of NTs grown in classic electrolytes (e.g. aqueous or organic), an organized spacing in between the nanotubes is clearly apparent. At the top, the average spacing is 109nm±46.7nm. Most important, to verify a true spacing in between the tubes, nanotube layers were cracked-off close to their substrate i.e. directly at their base, see Figure 1c. From the SEM images, it is evident that also at the bottom, spacing is established; an average tube to tube distance of 106nm ± 25nm is obtained. In other words, bottom and top spacing are in a very similar range, reflecting an almost cylindrical outer tube shape.

A schematic representation of such spaced nanotubes is shown in Figure 1d defining the main morphological parameters that will be further evaluated (outer diameter, spacing, wall thickness).

In the present case, i.e. for an electrolyte composition of 4wt% HF and 1wt% $H_2O$ in DEG, a controlled variation over the intertube spacing can be achieved through the applied voltage; at 10V a spacing of ≈21nm is obtained, whereas at 40V it results as ≈168nm; at the same time the tube diameter increases from 45 to 200nm (Figure 2a-d). From these data, a defined self-organized spacing of $TiO_2$ NTs is obtained in a limited voltage range 10–40V; at lower voltages (e.g. 5V) only a thin porous oxide is obtained and at higher voltages (>40V) nanotubes do not grow uniformly over the surface but anodization results in bundled tubes or the formation of a sponge-like oxide layer. This observation of a "region of existence" for sapced nanotube formation resembles findings reported for classic tubes (hexagonal packed) grown in EG, where with increasing voltage also first a transition from a nanoporous to a nanotubular morphology and at high voltages a transition to sponge-like oxide was observed [5,19,20].

If the morphological data for outer diameter and spacings are evaluated, linear trendlines are obtained in the 10–40V voltage range; at the higher/lower end of the range, quite high errors for diameter and spacing occur, due to the strongly increasing inhomogeneity in morphology. A statistical evaluation of the dependence of number of nanotubes per unit area, as a function of voltage is plotted in Figure 2e showing that in accord with increased spacing and tube diameter, a decrease of overall nanotube density is obtained.

Valuable insights can also be gained from the current density–time (J-t) plots (Figure 2f). For voltages that lie within the "ordered spacing regime" (10–40V), profiles present the typical regions encountered in nanotubes growth, i.e. the initial region where J is decreasing exponentially due to coverage of the anodized surface with a thin compact oxide film [6,21,22]. This is followed by a mild increase of J (usually ascribed to surface area increase due to initial porosification) and the third region, where a steady-state value is established (significantly higher than zero) where tubes grow under steady-state conditions [6,12]. Remarkable is that the difference between tubes grown with different spacing (in the "ordered spacing regime") lies in the initial stages of anodization (inset, 0-20min). After extended anodization time virtually identical steady-state J values are reached (inset, 120-240min). This indicates that self-organization and spacing dimensions are established in initial stages. In the later steady-state, i.e. where an equilibrium between film formation and dissolution is established, the current densities for spaced nanotubes regime have similar values $\approx 1.4 mA/cm^2$ (240min anodization) i.e. just a steady growth (and dissolution) of the pre-established patterns occurs.

Figure 3a illustrates the role of another key factor significantly influencing formation of spaced $TiO_2$ NTs, i.e. water content in the electrolyte. With higher water content (e.g. 20wt% $H_2O$) the spacing in between nanotubes disappears and a close-packed NTs structure without any distinct spacing in between nanotubes is obtained (Figure 3a). Whereas, when anodizing at same voltage (30V) but with an optimal water content (1wt% $H_2O$), spaced nanotubes are clearly observed (Figure 3b). If the voltage is increased under "optimal water" content conditions, the formation of a sponge oxide layer is observed (Figure 3c), showing no clear sign of self-organized ordering. As was previously shown [5,19] for classic organic electrolytes (EG), there are not only voltage thresholds for morphology changes (compact–tube/spongy) but also water content limits that induce significant morphology changes. In the present case, in particular the transition from hexagonally ordered to spaced nanotubes is strongly influenced by the water in electrolyte. Moreover, by applying high voltages, there is a transition from spaced nanotubes to sponge-like layers, as illustrated in Figure 3d.

These findings can be exploited to control diameter and spacing independently. Figure 3e-f shows NTs where the intertubular spacing is varied from 83 to 180nm, while maintaining similar diameter (≈200nm), achieved by adjusting both key parameters, voltage and water content. This example shows the broad variation possibilities inherent in control of these parameters to obtain a desired morphology of $TiO_2$ nanotubes and their spacing.

We believe that tuning the morphology and spacing of $TiO_2$ NTs arrays is highly promising to enhance or tune tubes in any application requiring higher surface area of a specific nanotopography.

**Acknowledgements**

The authors acknowledge the ERC, the DFG, and the DFG "Engineering of Advanced Materials" cluster of excellence and DFG "funCOS" for financial support.

**Figures and figure captions**

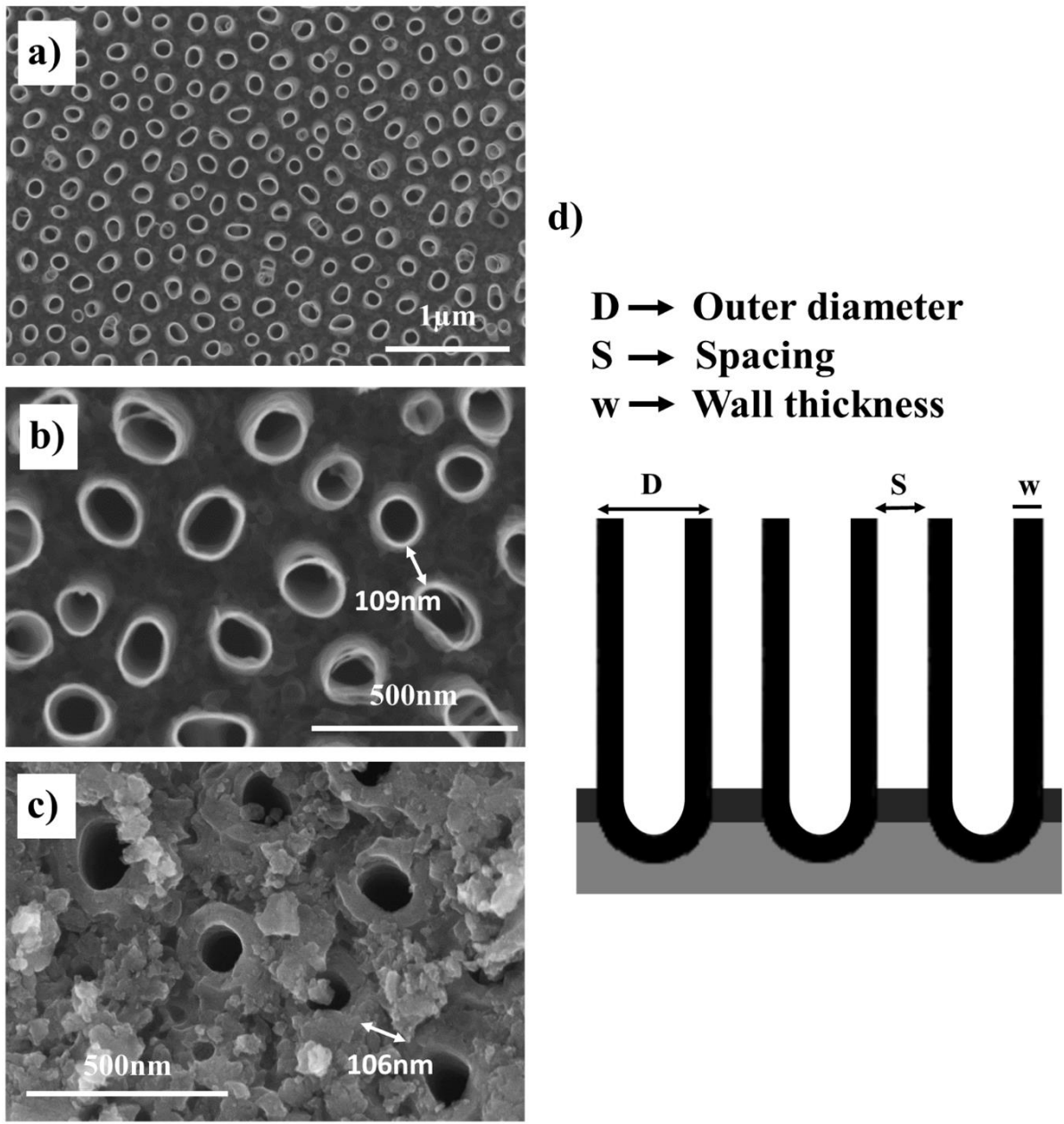

**Figure 1 Spaced TiO$_2$ nanotubes' morphology:** a) and b) top view SEM images of spaced TiO$_2$ nanotubes and c) top view image of nanotubes cracked close to the bottom. Nanotubes are obtained at 30 V in DEG + 4wt% HF + 0.3wt% NH$_4$F + 1wt% H$_2$O. d) Schematic of spaced TiO$_2$ nanotubes showing the main morphological characteristic.

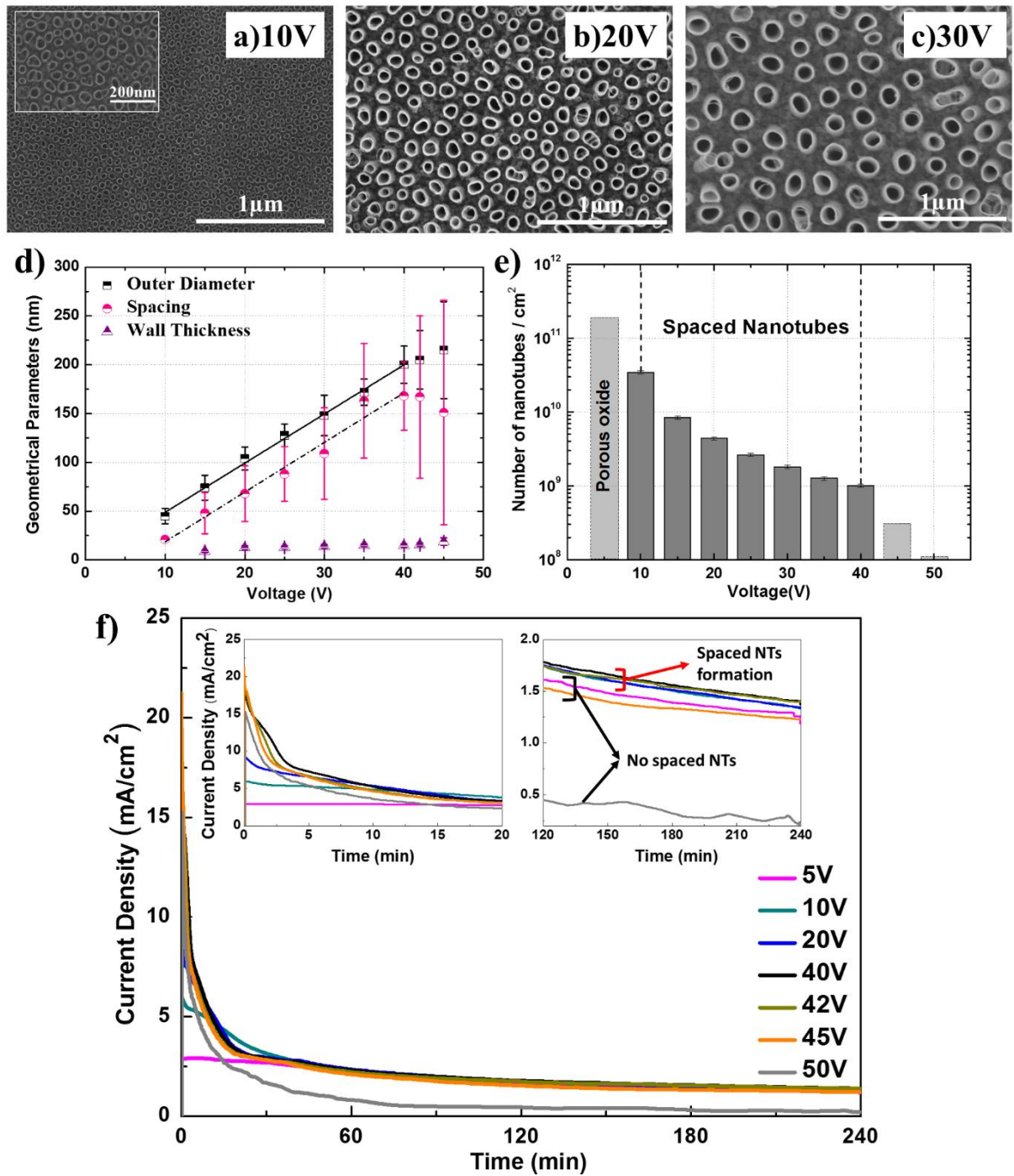

**Figure 2 Overview of spaced TiO$_2$ nanotubes characteristics:** Influence of the anodization applied voltage on the diameter and spacing of TiO$_2$ nanotubes, a) 10 V, b) 20 V and c) 30 V; d) Geometrical parameters (outer diameter, spacing, wall thickness) variation as a function of the applied voltage; e) Density of nanotubes as a function of the applied voltage and f) Selected current density − time profiles for anodizations at different voltages. Anodizations are performed in DEG + 4wt% HF + 0.3wt% NH$_4$F + 1wt% H$_2$O.

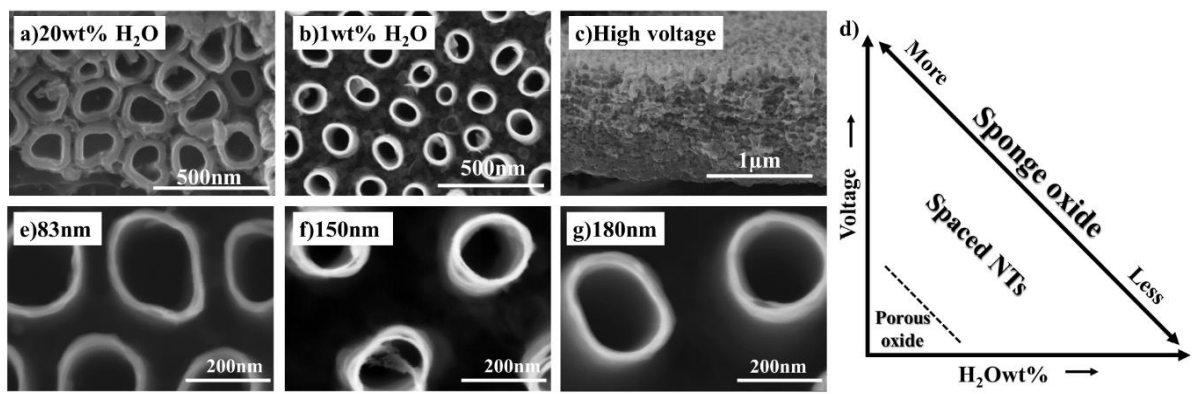

**Figure 3 Controlled morphology of spaced TiO$_2$ nanotubes:** Influence of water content on the morphology, a) top view SEM image of closed-packed TiO$_2$ nanotubes obtained in 20 wt% H$_2$O DEG based electrolyte, at an applied voltage of 30 V; and b) top view SEM image of spaced TiO$_2$ nanotubes obtained in similar anodization condition (30V) in 1 wt% H$_2$O DEG based electrolyte. d) Overview of the influence voltage and water content on the transition from pores to spaced nanotubes and to spongy oxide layers; Example of controlled spacing for TiO$_2$ nanotubes with a spacing of e) 83 nm, f) 150 nm and g) 180nm.